\documentclass[12pt]{article}
 \pdfoutput=1
 \usepackage{graphicx}
  \usepackage{epsfig}
  \usepackage{amsmath}
  \usepackage{amssymb}
  \usepackage{color}
  \newlength{\abstractwidth}
  \newcommand{\be}{\begin{equation}}
  \newcommand{\ee}{\end{equation}}
  
  \renewcommand{\title}[1]{\vbox{\center\bf{\Large{#1}}}\vspace{5mm}}
  \renewcommand{\author}[1]{\vbox{\center#1}\vspace{5mm}}
  \newcommand{\address}[1]{\vbox{\center\em#1}}
  \newcommand{\email}[1]{\vbox{\center\tt#1}\vspace{5mm}}

\begin{document}

\begin{titlepage}
\begin{center}
\hfill \\
\hfill \\
\vskip 1cm

\title{Measurements without Probabilities in the Final State Proposal}

\author{Raphael Bousso$^{1,2}$ and Douglas Stanford$^3$}

\address{$^1$Center for Theoretical Physics and Department of Physics,\\
 University of California, Berkeley, CA 94720, U.S.A.}
\address{$^2$Lawrence Berkeley National Laboratory, Berkeley, CA 94720,
  U.S.A.}
\address{$^3$Stanford Institute for Theoretical Physics and Department of Physics,\\ Stanford University, Stanford, CA 94305, U.S.A.}

\email{bousso@lbl.gov, salguod@stanford.edu}

\end{center}
  
  \begin{abstract}
The black hole final state proposal reconciles the infalling vacuum with the unitarity of the Hawking radiation, but only for some experiments. We study experiments that first verify the exterior, then the interior purification of the same Hawking particle. (This is the same protocol that renders the firewall paradox operationally meaningful in standard quantum mechanics.) We show that the decoherence functional fails to be diagonal, even upon inclusion of external ``pointer'' systems. Hence, probabilities for outcomes of these measurements are not defined. We conclude that the final state proposal does not offer a consistent alternative to the firewall hypothesis.
  \end{abstract}

\end{titlepage}

\tableofcontents

\baselineskip=17.63pt

\section{Introduction}
A key problem in quantum gravity is to understand how information emerges from evaporating black holes. Semiclassical dynamics requires that infalling observers find the vacuum state at the horizon; this leads to information loss \cite{hawking}. Black hole complementarity attempted to reconcile unitarity with the infalling vacuum by assigning different descriptions to exterior and infalling observers. But recently, Almheiri, Marolf, Polchinski and Sully (AMPS) \cite{amps} exhibited a conflict that arises in the description of the infalling observer alone. A late time Hawking mode $b$ is singled out when it is still near the horizon of a sufficiently old black hole. Unitarity requires $b$ to be nearly pure either alone or together with some subsystem $r_b$ of the early Hawking radiation. Smoothness of the horizon requires $b$ to be pure and highly entangled with a mode $\tilde{b}$ in the interior. Together, these requirements violate the strong subadditivity of the entanglement entropy, in the theory of the infalling observer. Thus, one must either give up unitarity (as Hawking originally advocated), or allow a singular ``firewall'' at the horizon (as suggested by AMPS).

The ``final state'' model of unitary black hole evaporation was proposed by Horowitz and Maldacena (HM) \cite{hm} as a solution to Hawking's original paradox. The HM proposal invokes a generalization of quantum mechanics that postselects on a final state at the black hole singularity.
In final state quantum mechanics, probabilities for histories, such as the outcomes of one or more experiments, are given by the diagonal entries of a decoherence functional, defined below. If the decoherence functional contains off-diagonal entries, then the set of histories fails to decohere and probabilities are not well-defined. 

For the HM proposal to succeed, it must resolve the conflict between unitarity and smoothness exhibited by Hawking and sharpened by AMPS. Intriguingly, strong subadditivity can be transcended in final state quantum mechanics. By choosing an initial state with $b,\tilde{b}$ in the vacuum and an appropriate final state, one can thus reconcile the demands of unitarity and smoothness \cite{kitaevpreskill,lloydpreskill}, in the following limited sense: an experiment that verifies whether $b$ and $r_b$ are in the correct entangled pure state (controlled by unitarity) will succeed with probability one; yet, so will an experiment that verifies whether $b$ and $\tilde{b}$ form the correct entangled pure state (the vacuum). Even better, if an experimenter first verifies the $b,\tilde{b}$ vacuum state and then the $b,r_b$ unitary state, both experiments are certain to succeed.

In this paper, we show that the proposal falls short if the experiments are performed in the opposite ordering. We find that the associated histories of outcomes fail to decohere, so their probabilities cannot be defined. In standard quantum mechanics, the functional can be made to decohere over physical measurements by explicitly including the interactions with a ``pointer'' (an external environment or apparatus). But as noted by Gottesman and Preskill \cite{gp} (GP), if the final state is to accomplish unitarity, it must undo entangling interactions between matter and interior modes such as $\tilde b$. Since the measurement of $b\tilde b$ is a special case of such an interaction \cite{Bou12}, the inclusion of pointers fails to decohere its outcome.  Hence, probabilities are fundamentally ill-defined for some experiments in the interior of the black hole.\footnote{Pointers could diagonalize the decoherence functional if (at the expense of unitarity) one neglected to include the GP correction to the final state. However, one would then obtain negligibly small probabilities for the correct $b,r_b$ and $b,\tilde b$ states~\cite{kitaevpreskill}. Thus, unitarity and smoothness at the horizon would both fail. Moreover, this approach would violate causality, since the outcome of the first measurement depends on whether the second one is performed.}

\paragraph{Outline} In Section \ref{stwo}, we review the Horowitz-Maldacena proposal, along with the Gottes\-man-Preskill refinement of the final state. We also review the decoherence functional as a tool for assigning probabilities. In Section \ref{sthree} we consider the alternative histories corresponding to experiments that verify unitarity and the vacuum in this order. We show that they do not decohere even upon explicit inclusion of pointers. We discuss our results in Section~\ref{sfour}.

Some calculational techniques are summarized in Appendix \ref{b}. In Appendix \ref{a}, we show that the oddities of the final state are invisible to a bulk observer who has access only to a sufficiently small, typical subsystem of the radiation.  For simplicity, we focus on s-wave Hawking quanta, and we work in the approximation where these quanta are maximally entangled. We expect our conclusions to apply more generally to thermally entangled quanta, and to any mode near the horizon that is minable in the sense of \cite{boussodouble}.

\section{Final State Quantum Mechanics}\label{stwo}

\subsection{Horowitz-Maldacena Proposal}\label{sHM}

\begin{figure}[ht]
\begin{center}
\includegraphics[scale = .9]{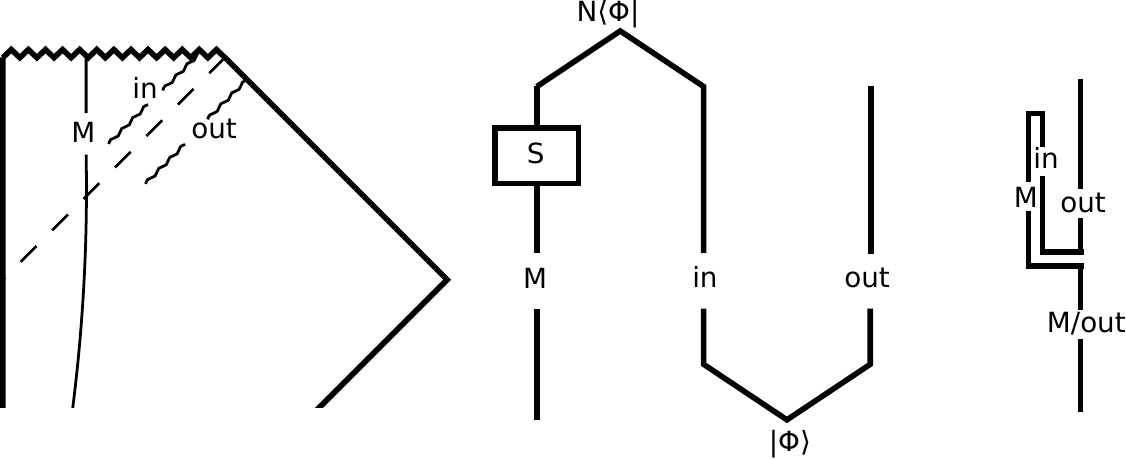}
\caption{The HM Hilbert spaces are defined on the left. The center shows the quantum circuit of the final state proposal \cite{gp}. In this diagram, diagonal lines meeting at a point represent the maximally entangled state $|\Phi\rangle$. If the lines open upwards, it is a ket vector, and if the lines open downwards, it is a bra. The drawing on the right is an interpretation in terms of path-integral folds similar to those discussed in \cite{hmps}.}\label{one}
\end{center}
\end{figure}

A black hole in asymptotically flat space forms and then evaporates into a cloud of Hawking radiation. HM associate three Hilbert spaces to this system, illustrated in the left panel of Fig.~\ref{one}. $M$ represents the Hilbert space of the matter that forms the black hole, $out$ represents the outgoing Hawking modes, and $in$ represents their Unruh partners. These subsystems are understood as separate tensor factors, each of dimension $N = e^{S_{BH}} = e^{A/4G_N}$. 

The goal of the HM proposal is to reconcile the unitary evaporation
\be
|m\rangle_M \rightarrow S_{jm}|j\rangle_{out}
\ee
with the semiclassical analysis of Hawking radiation, which requires that the $in$ and $out$ systems should be in the highly entangled Unruh state, modeled as
\begin{align}
|{\cal U}\rangle &= |\Phi\rangle_{in,out} \\
|\Phi\rangle_{X,Y}&\equiv \frac{1}{dim(X)^{1/2}}\sum_{i=1}^{dim(X)}|i\rangle_{X}|i\rangle_{Y}.
\end{align}
HM proceed by introducing the $in$ and $out$ systems in the infalling vacuum state $|{\cal U}\rangle$, and then ``projecting'' the $M$ and $in$ systems against the super-normalized final state 
\be \label{eq-final}
\langle BH| = N^{1/2}\sum_{m,i}S_{im}\langle m|_M\langle i|_{in}  = N \langle \Phi |_{M, in} (S\otimes 1).
\ee
The net effect of tensoring in the $in,out$ systems and then projecting against $\langle BH|$ is to map the initial state of the $M$ system as
\begin{align}
|m\rangle &\rightarrow |m\rangle|{\cal U}\rangle \notag\\
&\rightarrow \langle BH|\Big(|m\rangle|{\cal U}\rangle\Big) \\
& = S_{jm}|j\rangle \notag,
\end{align}
which is the desired state of the $out$ system. This complete evolution is illustrated as a quantum circuit in the center panel of Fig.~\ref{one}. {(See Ref.~\cite{eva} for string theory arguments supporting the proposal.)}

In order to exhibit the relation between AMPS and HM, we consider an old black hole that has emitted more than half of its initial entropy \cite{page2}. We focus on a Hawking quantum $b$ that is still in the near horizon zone, its interior partner $\tilde b$ with which $b$ forms the infalling vacuum, and a subsystem $r_b$ of the early Hawking radiation that purifies $b$ in the unitary out-state \cite{haydenpreskill}.\footnote{A physical Hawking particle will be thermally but not maximally entangled. Since arbitrary occupation numbers are involved, a full treatment would require Hilbert spaces of infinite dimension. The final state would be chosen with inverse Boltzmann factors that compensate for the thermal factors appearing in the Unruh state. This challenge is surmountable and unrelated to the obstruction we identify here. For simplicity of presentation, we will follow~\cite{lloydpreskill} in treating individual modes and their purifications as finite-dimensional maximally entangled systems.}

\begin{figure}[ht]
\begin{center}
\includegraphics[scale = 1.1]{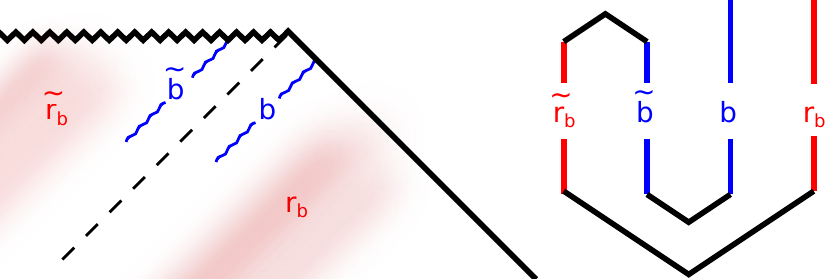}
\caption{If we focus on a particular state of the infalling matter, then the final state projects the interior partner of $b$ (called $\tilde{b}$) with the nonlocal interior partner of $r_b$ (called $\tilde{r_b}$).}\label{two}
\end{center}
\end{figure}

The definition of $r_b$ depends on the initial state of the black hole, and we will fix this to be some reference state $|0\rangle_M$. Given this initial state of the matter system, the final state projection reduces to a projection of the $in$ system with the bra vector $N^{1/2}\sum_i \langle i|S_{i0}$. Generically, this $in$ final state will involve maximal entanglement between $\tilde{b}$ and some other (highly nonlocal) subsystem of the $in$ Hilbert space, which we denote $\tilde{r_b}$. The Unruh partner of this complicated operator $\tilde{r_b}$ is denoted $r_b$, and becomes the subsystem of the early radiation that is entangled with $b$. This arrangement is represented in Fig.~\ref{two}.
With these definitions, $\tilde{r_b},\tilde{b},b,r_b$ are uncorrelated with the remainder of the system. Choosing suitable bases, we write their initial state as
\be
|\Phi\rangle_{\tilde{b},b}|\Phi\rangle_{\tilde{r_b},r_b}.
\ee
The final state for the $\tilde{b},\tilde{r_b}$ subsystem is
\be
d\langle \Phi|_{\tilde{r_b},\tilde{b}}.
\ee
Here, $d$ is the Hilbert space dimension of each factor, e.g.\ of $b$.

\subsection{Gottesman-Preskill Refinement}

\begin{figure}[ht]
\begin{center}
\includegraphics[scale = .9]{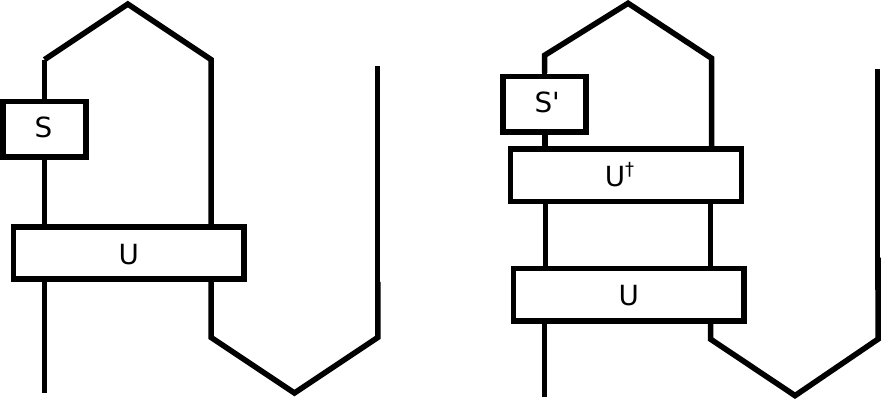}
\caption{The Gottesman-Preskill refinement \cite{gp}: the diagram at left does not provide a unitary map from $M$ to $out$, due to interactions $U$ between the matter and the in modes. In the right diagram, we add a compensating $U^\dagger$ to the final state. This ``undoes'' the interaction behind the horizon, up to a modification of the $S$ matrix to $S' = SV$, and results in a unitary circuit.}\label{three}
\end{center}
\end{figure}

So far, we have treated the $M$, $in$ and $out$ systems as noninteracting. In this approximation, unitarity of the map from $M$ to $out$ is equivalent to the requirement that $in$ and $M$ be maximally entangled in the final state. However, the experiments that we will analyze in \S\ref{sthree} involve interaction between $M$ and $in$. Preserving unitarity in the presence of such interactions requires an adjustment of the final state, as emphasized by Gottesman and Preskill \cite{gp}. Let us represent the bulk time evolution from the horizon to the singularity as a unitary operator $U$. From the perspective of the quantum circuit, drawn in the left panel of Fig.~\ref{three}, this interaction effectively modifies the final state to $\langle BH|U$, which will not in general be maximally entangled. To restore unitarity, we must replace Eq.~(\ref{eq-final}) with a final state of the form
\be \label{eq-final2}
\langle BH|(V_M\otimes 1_{in})U^\dagger~,
\ee 
with $V$ a unitary that acts only on the $M$ system, and which we include for the sake of generality. This $V$ can be absorbed into a modification of the $S$ matrix to $S' = SV$.

The presence of the $U^\dagger$ will be crucial for our analysis in \S\ref{sthree}. We might interpret it as ``undoing'' the dynamics behind the horizon prior to projection. Another way to think about it is that the final state is really imposed at the horizon, and the interior is a path integral fold, a representation of the unit operator as $U U^\dagger$. This interpretation is sketched in the right panel of Fig.~\ref{one}.

\subsection{Decoherence Functional and Probabilities}

The assignment of probabilities to measurement outcomes is subtle in the HM setup. This is partially because of the final state projection, and partially because the interactions that are normally implicit in a quantum measurement have to be accounted for in the definition of the final state. Both of these complications can be treated carefully using the decoherence functional formalism, developed by Gell-Mann and Hartle \cite{gellmannhartle}, studied further in \cite{dowkerhalliwell,gellmannhartle2}, and applied to the final state proposal in the original paper of Horowitz and Maldacena. Here we will give a general summary, before applying the formalism to the black hole final state in the next section.

The decoherence functional depends on a pair of histories, $\alpha_1,...,\alpha_n$ and $\alpha_1',...,\alpha_n'$, represented by a sequence of projection operators $\Pi_{\alpha_n}\Pi_{\alpha_{n-1}}...\Pi_{\alpha_1}$ and $\Pi_{\alpha_n'}\Pi_{\alpha_{n-1}'}...\Pi_{\alpha_1'}$. It is defined as
\begin{align}
D(\alpha,\alpha') = tr[\sigma C_\alpha \rho C_\alpha'^\dagger] \\
C_\alpha = \Pi_{\alpha_n}\Pi_{\alpha_{n-1}}...\Pi_{\alpha_1}.
\end{align}
Here, $\rho$ is the density matrix of the system, and $\sigma$ is the matrix that represents the final state, normalized so that $tr[\sigma\rho]=1$.

When the decoherence functional is diagonal, one can assign a probability to a history by taking the normalized diagonal entries
\be \label{eq-decfun}
P(\alpha) = D(\alpha,\alpha).
\ee
Diagonality ensures\footnote{Some authors impose a weaker condition that the off-diagonal entries should be purely imaginary. Here, we show explicitly that joint probabilities do not sum to the individual probabilities, so the distinction is moot.} that the resulting quantities actually have the properties of probabilities, e.g.
\be
\sum_\beta P(\alpha,\beta) = P(\alpha).
\ee
When the decoherence functional is not diagonal, probabilities cannot be assigned to the histories. 

If $\sigma = 1$, this is an equivalent description of standard quantum mechanics. For any one-step history ($n = 1$), $D$ is diagonal and the probabilities are simply $tr[\Pi_\alpha \rho]$. However, for histories involving more than one step, $D$ may not be diagonal. As a simple example, consider a qubit that begins with a definite $x$ spin, $\rho = |+\rangle\langle+|$. If we consider histories that begin with this state, have definite values of the $z$ spin, and then definite values of the $x$ spin, we find that the decoherence functional is not diagonal. Probabilities for the $z$ spin at the intermediate time cannot be assigned.

Of course, if we {\it measure} the $z$ spin, we can assign probabilities. The decoherence functional {applies to a closed system and so does not incorporate measurement by external systems}. To describe actual measurements, {we need to include an apparatus or environment, and its dynamical interactions.} This can be done by introducing pointers that couple to the system, {one pointer for each measurement.} For example, measurement of the first variable $\alpha_1$ by a pointer initialized in state $|0\rangle_1$ involves the evolution
\be
\label{p1}
|\Psi\rangle|0\rangle_1 \rightarrow \sum_{\alpha_1} \Big(\Pi_{\alpha_1}|\Psi\rangle\Big)|\alpha_1\rangle_1.
\ee
Measurement of the second variable then corresponds to
\be
\label{p2}
\sum_{\alpha_1} \Big(\Pi_{\alpha_1}|\Psi\rangle\Big)|\alpha_1\rangle_1|0\rangle_2 \rightarrow \sum_{\alpha_1,\alpha_2}\Big(\Pi_{\alpha_2}\Pi_{\alpha_1}|\Psi\rangle\Big)|\alpha_1\rangle_1|\alpha_2\rangle_2.
\ee
The effect after $n$ such measurements is to prepare a state in which each history of projectors $\Pi_{\alpha_n}...\Pi_{\alpha_1}$ is paired with a distinct element $|\alpha_1\rangle_1...|\alpha_n\rangle_n$ of an orthonormal basis of pointer states. If the final state of the system is trivial, i.e. $\sigma = 1$, the entire effect is to set all off-diagonal elements of $D$ to zero. However, if the final state acts nontrivially on the pointers, off-diagonal terms can remain. This will be important in what follows.

\section{Measurements without Probabilities} \label{sthree}

In this section, we will use the decoherence functional to study the probabilities of histories that verify the double entanglement of $b,\tilde{b}$ and $b,r_b$. The verification of $b,r_b$ entanglement is associated to a resolution of the identity by $\Pi_{r_b\,b}$ and $(1 - \Pi_{r_b\,b})$, where 
\be
\Pi_{XY} = |\Phi\rangle\langle\Phi|_{XY} \otimes 1
\ee
is the projector onto the maximally entangled state $|\Phi\rangle$ of $XY$, tensored with the identity on the rest of the system. Similarly, the verification of $\tilde{b},b$ entanglement is associated to the projectors $\Pi_{\tilde{b}\,b}$ and $(1 - \Pi_{\tilde{b}\,b})$.

In Sec.~\ref{sec-orig}, we will consider the original Horowitz-Maldacena proposal. We will show that it does not yield well-defined probabilities for certain experiments that check both entanglements. In Sec.~\ref{sec-point}, we show that the interaction with an environment does not resolve this problem, because the Gottesman-Preskill objection can only be resolved by undoing such interactions and so robbing them of their decoherent effect.

\subsection{Verification of Double Entanglement Fails to Decohere}
\label{sec-orig}

In the application of Eq.~(\ref{eq-decfun}) to the Horowitz-Maldacena proposal, we have
\begin{align}
\sigma &= |BH\rangle\langle BH|_{M,in}\otimes 1_{out} \label{final}\\
\rho &= \rho_M \otimes |{\cal U}\rangle\langle {\cal U}|_{in,out} \label{initial}.
\end{align}
Assuming a typical initial state of the matter system, we can define the $\tilde{r_b},\tilde{b},b,r_b$ systems as in \S\ref{sHM}. The initial and final states of these subsystems are
\begin{align}
\sigma &= d^2|\Phi\rangle\langle\Phi|_{\tilde{r_b},\tilde{b}}\otimes 1_{b,r_b} \label{finals}\\
\rho &= |\Phi\rangle\langle \Phi|_{\tilde{r_b},r_b}\otimes |\Phi\rangle\langle \Phi|_{\tilde{b},b}.\label{initials}
\end{align}
We do not encounter any subtleties in assigning probabilities to histories with definite outcomes for one or the other of the $\tilde{b},b$ and $b,r_b$ tests. In fact, the decoherence functional is diagonal without any need for measurement by pointers, and it indicates that the probability of correct entanglement is one. It is also straightforward to compute the decoherence functional for two-event histories in which $\tilde{b},b$ are definitely in (or not in) state $|\Phi\rangle_{\tilde{b}\,b}$, and then $b,r_b$ are definitely in (or not in) state $|\Phi\rangle_{r_b\,b}$. One finds that the only nonzero element of $D$ is the diagonal element corresponding to successful verification of both entanglements.

We will therefore focus attention on the opposite ordering, in which we try to assign a probability to a history with a definite $b,r_b$ result followed by a definite $\tilde{b},b$ result. In this case, the decoherence functional is not diagonal. The explicit values are simple to compute using the diagrams discussed Appendix \ref{b}, and the full matrix of results is shown in Table~\ref{table}. 
\begin{center}
\begin{table}[ht]
    \begin{tabular}{c|cccc}
    ~                                        & $\Pi_{r_b\,b}\Pi_{\tilde{b}\,b}$ & $\Pi_{r_b\,b}(1-\Pi_{\tilde{b}\,b})$ & $(1-\Pi_{r_b\,b})\Pi_{\tilde{b}\,b}$ & $(1-\Pi_{r_b\,b})(1-\Pi_{\tilde{b}\,b})$ \\ \hline
    $\Pi_{\tilde{b}\,b}\Pi_{r_b\,b}$         & $\frac{1}{d^4}$                  & $\frac{1}{d^2} - \frac{1}{d^4}$      & $\frac{1}{d^2} - \frac{1}{d^4}$      & $\frac{1}{d^4} - \frac{1}{d^2}$          \\
    $(1-\Pi_{\tilde{b}\,b})\Pi_{r_b\,b}$     & $\frac{1}{d^2} - \frac{1}{d^4}$  & $\left(1 - \frac{1}{d^2}\right)^2$   & $\left(1 - \frac{1}{d^2}\right)^2$   & $-\left(1 - \frac{1}{d^2}\right)^2$      \\
    $\Pi_{\tilde{b}\,b}(1-\Pi_{r_b\,b})$     & $\frac{1}{d^2} - \frac{1}{d^4}$  & $\left(1 - \frac{1}{d^2}\right)^2$   & $\left(1 - \frac{1}{d^2}\right)^2$   & $-\left(1 - \frac{1}{d^2}\right)^2$      \\
    $(1-\Pi_{\tilde{b}\,b})(1-\Pi_{r_b\,b})$ & $\frac{1}{d^4} - \frac{1}{d^2}$  & $-\left(1 - \frac{1}{d^2}\right)^2$  & $-\left(1 - \frac{1}{d^2}\right)^2$  & $\left(1 - \frac{1}{d^2}\right)^2$       \\
    \end{tabular}
\caption{The decoherence functional $D(\alpha,\alpha')$ for the histories associated to a definite $b,r_b$ result followed by a definite $\tilde{b},b$ result, without any coupling to pointers. $d$ is the Hilbert space dimension of each subsystem. The rows are values of $C_{\alpha}$ and the columns are values of $C_{\alpha'}^\dagger$. The corresponding table for the opposite ordering of the projectors would have a one in the top left corner and zeros elsewhere.}\label{table}
\end{table}
\end{center}
The non-diagonality of $D$ means that the system does not naturally decohere into histories with definite outcomes for the $b,r_b$ and $\tilde{b},b$ measurements, performed in that order.

\subsection{Pointers Fail to Enforce Decoherence Inside the Horizon}
\label{sec-point}

In standard quantum mechanics, we could force the histories to decohere by adding pointers to the system and coupling them to $b,r_b$ and $\tilde{b},b$ as in Eqs.~(\ref{p1}) and (\ref{p2}). This procedure is straightforward for the $b,r_b$ measurement, which can be performed outside the horizon. Hence, the pointer can be taken to remain outside the horizon, so it will not be constrained by the final state. Tracing over the pointer then forces the decoherence functional to be diagonal in the $b,r_b$ history, as in Table~\ref{table2}.\footnote{In what follows we will suppress the explicit pointer that interacts with $b,r_b$, and simply restrict to diagonal terms for this measurement.}

We could try the same procedure for the $\tilde{b},b$ measurement, naively coupling to pointers with a trivial final state, and thus setting the remaining off-diagonal entries to zero. This prescription immediately runs into trouble. To begin with, the diagonal elements of Table~\ref{table} do not sum to one. In the standard treatment of postselected quantum mechanics, this would not be cause for alarm; one simply adjusts the normalization. However, if we do this, we find that the chance for the initial $b,r_b$ measurement to verify the correct entanglement is a little over a third, compared to a probability of one in the case where no later $\tilde{b},b$ measurement is made.
\begin{center}
\begin{table}[ht]
    \begin{tabular}{c|cccc}
    ~                                        & $\Pi_{r_b\,b}\Pi_{\tilde{b}\,b}$ & $\Pi_{r_b\,b}(1-\Pi_{\tilde{b}\,b})$ & $(1-\Pi_{r_b\,b})\Pi_{\tilde{b}\,b}$ & $(1-\Pi_{r_b\,b})(1-\Pi_{\tilde{b}\,b})$ \\ \hline
    $\Pi_{\tilde{b}\,b}\Pi_{r_b\,b}$         & $\frac{1}{d^4}$                  & $\frac{1}{d^2} - \frac{1}{d^4}$      & 0                                    & 0                                        \\
    $(1-\Pi_{\tilde{b}\,b})\Pi_{r_b\,b}$     & $\frac{1}{d^2} - \frac{1}{d^4}$  & $\left(1 - \frac{1}{d^2}\right)^2$   & 0                                    & 0                                        \\
    $\Pi_{\tilde{b}\,b}(1-\Pi_{r_b\,b})$     & 0                                & 0                                    & $\left(1 - \frac{1}{d^2}\right)^2$   & $-\left(1 - \frac{1}{d^2}\right)^2$      \\
    $(1-\Pi_{\tilde{b}\,b})(1-\Pi_{r_b\,b})$ & 0                                & 0                                    & $-\left(1 - \frac{1}{d^2}\right)^2$  & $\left(1 - \frac{1}{d^2}\right)^2$       \\
    \end{tabular}
\caption{If we couple to pointers, we can decohere the $b,r_b$ measurement, but $D(\alpha,\alpha')$ remains off-diagonal in the outcome of the $\tilde{b},b$ measurement.}\label{table2}
\end{table}
\end{center}

This violation of causality can be traced to the fact that we are not correctly treating the interaction between $\tilde b,b$ and the pointer. This interaction must take place behind the horizon, and is therefore a special case of the $U$ interaction discussed in Section~\ref{stwo}. The GP refinement requires us to adjust the final state with a compensating $VU^\dagger$. Thus, we compute
\be
D(\alpha,\alpha') = tr[UV^\dagger\sigma VU^\dagger \Pi_{\alpha_2}U\Pi_{\alpha_1}\rho \Pi_{\alpha_1}U^\dagger\Pi_{\alpha_2'}].
\ee
Here, $\Pi_{\alpha_2}$ and $\Pi_{\alpha_2'}$ run over $\Pi_{\tilde{b}\,b}$ and $(1-\Pi_{\tilde{b}\,b})$, while $\Pi_{\alpha_1}$ is either $\Pi_{b\,r_b}$ or $(1 - \Pi_{b\,r_b})$. We have enlarged the Hilbert space to include a pointer with initial state $|0\rangle\langle 0|$. $U$ is the unitary that entangles the pointer with $\tilde{b},b$:
\be
U|\Psi\rangle|0\rangle = \Pi_{\tilde{b}\,b}|\Psi\rangle|0\rangle + (1-\Pi_{\tilde{b}\,b})|\Psi\rangle|1\rangle.
\ee

We will now argue that this decoherence functional is the same as the one without any $U$ or $V$ matrices included, i.e. that consistently including the pointer is equivalent to not including it at all. Notice that the effect of including $U$ and $V$ is to replace the original $\Pi_{\tilde{b}\,b}$ operator with $U^\dagger \Pi_{\tilde{b}\,b}U$ and to update the $S$ matrix to $S' = SV$.\footnote{The $S$ matrix now acts on a larger Hilbert space, including the pointer that falls into the black hole.} Since
\be
U^\dagger \Pi_{\tilde{b}\,b} U |\Psi\rangle|0\rangle = \Pi_{\tilde{b}\, b} |\Psi\rangle|0\rangle,
\ee
it is clear that the $U$ matrix has no effect. In general, changing the $S$ matrix will influence the decoherence functional. However, if we wish to preserve causality, then throwing a pointer into the black hole should not affect the quantum state of the $b,r_b$ subsystem that has already emerged. This means that $V$ must not affect the final state for the $\tilde{r_b},\tilde{b}$ subsystem. Since this is the only part of the final state that is relevant for our analysis, we can set $V$ to one.

Our conclusion is that pointers don't decohere $\tilde{b},b$. Unitarity requires the final state to undo measurements of the $in$ modes up to a transformation $V$ that is restricted by causality to have no effect. We are stuck with the non-diagonal decoherence functional from Table~\ref{table2}, and are therefore unable to assign probabilities.

\section{Discussion} \label{sfour}

In this paper, we pointed out a difficulty with the final state projection model of black hole evaporation. This difficulty appears to be distinct from the firewall, but arises from a careful consideration of the same subsystems relevant for the AMPS argument. In ordinary quantum mechanics, the simultaneous demands of unitarity and the infalling vacuum simply cannot be satisfied by these systems~\cite{amps}. The attempt to reconcile unitarity with the infalling vacuum by appealing to final state quantum mechanics renders ill-defined the probabilities for outcomes of certain measurements involving the same subsystems. (It is interesting that several other, apparently distinct attempts to evade firewalls encounter the same problem, that the horizon remains special to a local observer~\cite{boussodouble,boussofrozen}.)

It would be nice to find a modification or extension of the final state proposal that evades the obstruction. It is tempting to speculate that computational complexity arguments provide the needed patch. The experiments with ill-defined probabilities involve the subsystem $r_b$, which Harlow and Hayden \cite{harlowhayden} have argued requires a time exponential in $S_{BH}$ to extract -- far longer than the lifetime of an evaporating black hole. This suggests an out: the theory need not provide probabilities for the experiments from Section \ref{sthree}, because such experiments cannot be performed. However, the computational obstacles of \cite{harlowhayden} appear to be surmountable in the ``laboratory'' setting of a large black hole in an AdS box \cite{AMPSS}. Moreover, there exist Haar-rare states for the black hole-radiation system where the extraction of $r_b$ is fast; yet, firewalls in these states would appear to be no more acceptable than for typical black holes~\cite{boussodouble}.

It is worth stressing that we cannot fall back on what may be a more familiar treatment of decoherent measurements. In ordinary quantum mechanics with $\sigma=1$, the computation of probabilities as traces with appropriate projection operators, via Eq.~(\ref{eq-decfun}), is merely optional. Instead, given the complete (entangled) quantum state of a system and any apparatus or environment it has interacted with, one can perform a partial trace over the environment and obtain a density operator for the system, whose eigenvalues can be interpreted as the probabilities for the corresponding outcomes. But in final state quantum mechanics, there is no unique well-defined quantum state, since the initial and final state are treated symmetrically. In this setting, the decoherence functional provides the {\em only} prescription for obtaining probabilities; and here it fails to diagonalize even if pointers are included.

One could contemplate devising rules for computing probabilities that depend on whether a measurement is performed inside or outside the black hole. We are not aware of a specific proposal. As general constraint, one would have to ensure that despite the differing rules, the probabilities themselves are not sharply sensitive to the location. Otherwise, a clever observer could exploit this to detect the horizon by local experiments. This difficulty resembles the ``frozen vacuum'' objection to the nonunitary identification of Hilbert spaces (``ER$=$EPR'', ``$A=R_B$'')~\cite{boussofrozen}. If the horizon is vacuous but distinct from other vacuum regions on scales below the curvature scale, the equivalence principle is violated. 

\paragraph{Acknowledgments}  We thank A.~Kitaev, J.~Preskill and V.~Rosenhaus for discussions. The work of R.B.\ is supported by the Berkeley Center for Theoretical Physics, by the National Science Foundation (grant number 1214644), by the Foundational Questions Institute, by ``New Frontiers in Astronomy and Cosmology'', and by the U.S.\ Department of Energy (DE-AC02-05CH11231). The work of D.S.\ is supported by the Stanford Institute for Theoretical Physics and NSF Grant 0756174. We both acknowledge the hospitality of the Kavli Institute for Theoretical Physics, supported by NSF Grant PHY11-25915.

\begin{appendix}

\section{Diagrams for the Decoherence Functional}\label{b}
\begin{figure}[ht]
\begin{center}
\includegraphics[scale = .35]{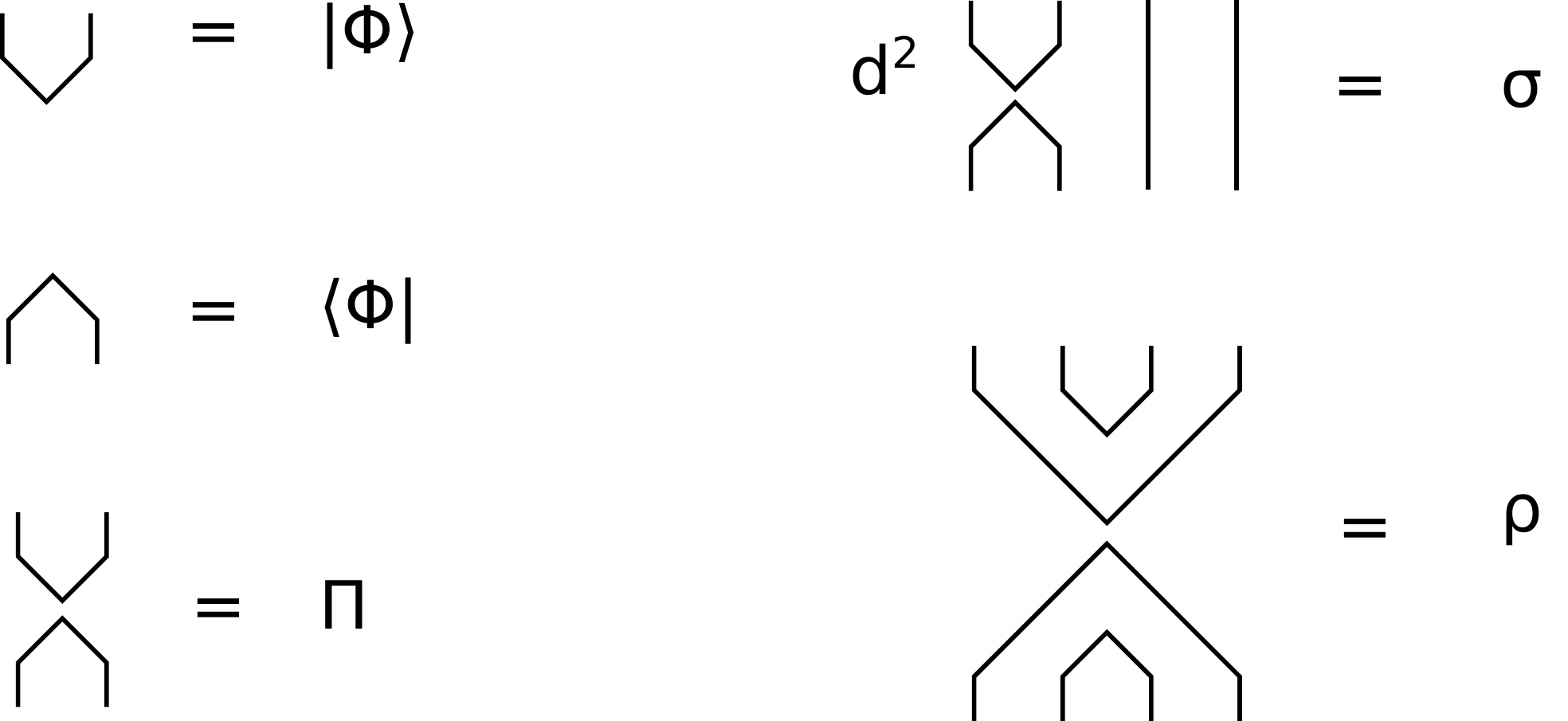}
\caption{Index diagrams as described in the text.}\label{rules}
\end{center}
\end{figure}

In Section~\ref{sthree}, we wrote down the decoherence functional for histories involving $\tilde{b},b$ and $r_b$. In this appendix, we will illustrate a diagrammatic approach for computing $D$.\footnote{We thank John Preskill for teaching us how to use these diagrams.} The idea is to use a simple form of quantum circuit, in which endpoints of lines represent indices of a state vector, and lines represent Kronecker-delta propagators. With this convention, we can represent the particular maximally entangled ket vector $|\Phi\rangle = d^{-1/2}\sum_i |i\rangle|i\rangle$ as a ``cup,'' and the corresponding bra as a ``cap.'' This is illustrated in Fig.~\ref{rules}. A cup has two uncontracted indices, but they are constrained to be equal. This encodes the state $|\Phi\rangle$, provided that we remember to associate a factor of $d^{-1/2}$.

To represent a matrix, we use the outer product of ket and bra vectors. The top ends of lines represent the left indices of the matrix, and the bottom ends represent the right indices. Thus a vertical line segment represents the identity matrix $\delta_{ab}$, while a cup sitting on top of a cap represents the projector $\Pi = |\Phi\rangle\langle\Phi|$. Multiple upper or lower indices are understood in the sense of a tensor product. 

\begin{figure}[ht]
\begin{center}
\includegraphics[scale = .35]{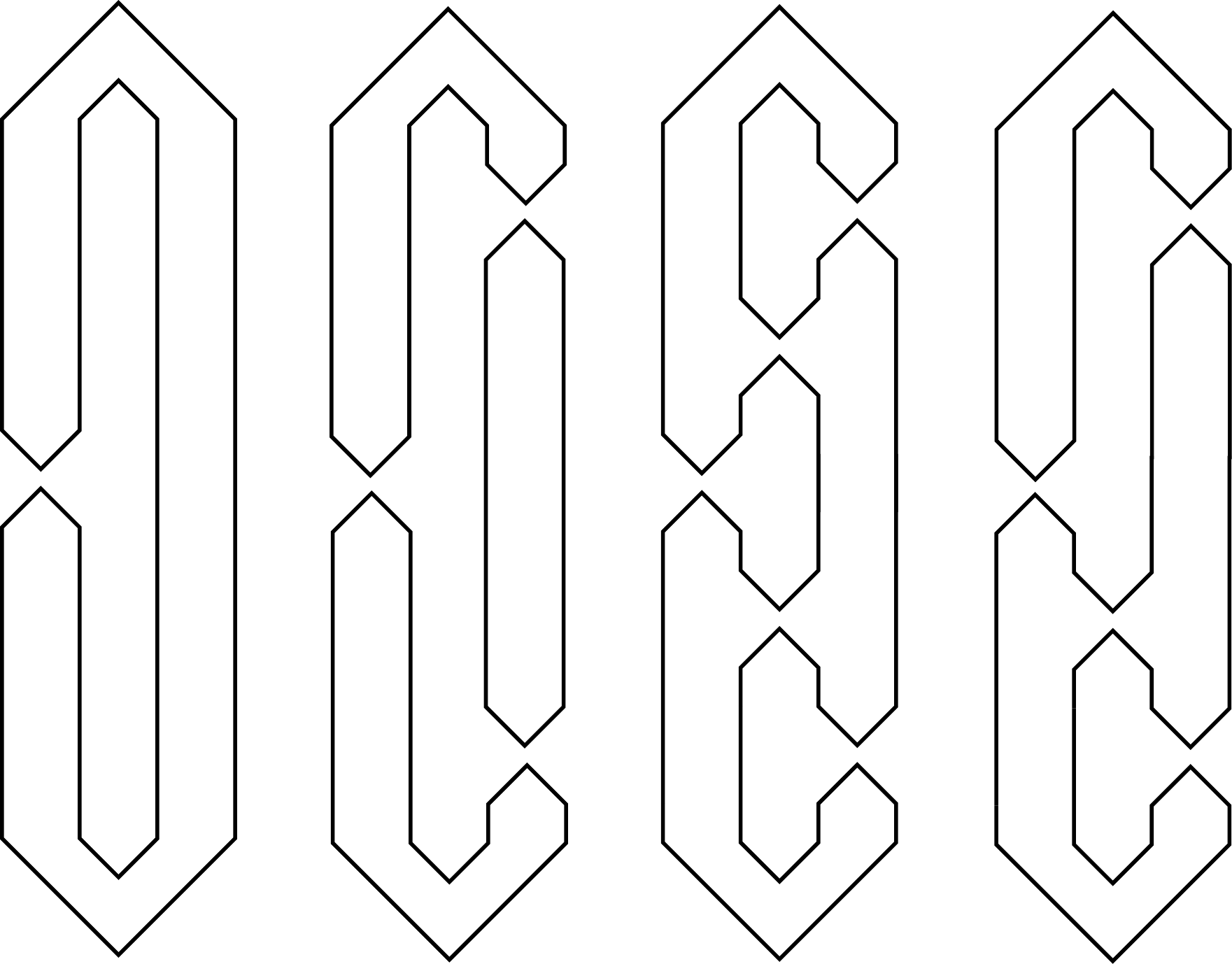}
\caption{Diagrams for four different traces discussed in the main text.}\label{loops}
\end{center}
\end{figure}

With these ingredients, we can build diagrams for the initial and final matrices of the $\tilde{r_b},\tilde{b},b,r_b$ subsystems, Eq.s (\ref{initials}) and (\ref{finals}). The diagrams are shown in Fig.~\ref{rules}, with the convention that the free indices (from left to right) are associated to the tensor factors $\tilde{r_b},\tilde{b},b,r_b$. The product $\sigma\rho$, is obtained by contracting the lower indices of $\sigma$ with the upper indices of $\rho$. The trace is then given by contracting the lower indices of $\rho$ with the upper indices of $\sigma$, leading to the diagram at left in Fig.~\ref{loops}. This diagram has one connected loop. Summing the index in this loop gives a factor of $d$. We also have six ``cups'' or ``caps,'' one for each copy of $|\Phi\rangle$ or $\langle\Phi|$. Each one comes with a factor of $d^{-1/2}$. Together with the loop, we have $d^{-2}$ for the diagram, which is cancelled by the explicit $d^2$ in the normalization of $\sigma$.

For the purposes of Section~\ref{sthree}, we need to compute the trace with projectors inserted. These can be added using the representation of $\Pi$ as a ``cup-cap.'' A few examples are worked out in Fig.~\ref{loops}. The second diagram from the left computes $tr[\sigma \Pi_{b,r_b}\rho \Pi_{b,r_b}]$. We have ten cups or caps, three loops, and an overall $d^2$ for the normalization of $\sigma$, giving one. The third diagram from the left, for $tr[\sigma\Pi_{\tilde{b},b}\Pi_{b,r_b}\rho \Pi_{b,r_b}\Pi_{\tilde{b},b}]$ has fourteen cups or caps and only one loop, for a total of $1/d^4$. The final diagram computes $tr[\sigma\Pi_{\tilde{b},b}\Pi_{b,r_b}\rho \Pi_{b,r_b}]$ as $1/d^2$.

\section{Simple Subsystems Cannot Detect the Final State}\label{a}
In this appendix, we will show that if the $S$ matrix is sufficiently random\footnote{In our computations, we will assume a Haar random $S$, but a matrix drawn from a unitary 2-design would give equivalent results.}, then measurements restricted to a fifth of each of the $M$, $in$, and $out$ systems will not be sensitive to the final state. More precisely, let us define subsystems $M_1,in_1,out_1$ of equal dimension $n$, and tensor complements $M_2,in_2,out_2$ each of dimension $N/n$. In this decomposition, $in_1$ and $out_1$ are the measured $in$ and $out$ modes plus their Unruh partners, and the $M$ system represents an appratus that falls into the black hole. Aside from the restriction of Unruh partnership, the subsystems are arbitrary but must be chosen independently of $S$.  Finally, we will also restrict the state of the rest of the matter system, $M_2$ to be $|0\rangle$. We will prove that, averaging over $S$ matrices with the Haar measure, the expected difference in the decoherence functional 
\be
\int dS \, \Big|tr[\sigma(S) \Pi\rho\Pi'] - tr[\Pi\rho\Pi']\Big| \le \sqrt{\frac{n^5}{N}}
\label{answer}
\ee
is small for any projectors $\Pi,\Pi'$ that act as the identity on the ``2'' subsystems. This implies that if each of the ``1'' subsystems are smaller than a fifth of their respective systems, no projective measurement can detect the final state with significant probability. We expect that the one-fifth bound is suboptimal. It is very likely that an improvement on our argument could show the final state cannot be detected as long as the subsystems are smaller than a third, or possibly even a half.\footnote{Even if we ignore $M_1$ and $in_1$, an observer with access to more than half of $out$, must be able to detect the final state, since they can detect that the Hawking radiation is not maximally mixed \cite{page}.}

Intuitively, the reason is that if we restrict the final matrix $\sigma$ to a small enough subsystem, it will closely approximate the identity matrix. For a general final state, this would not be sufficient, since the partial trace of a product of matrices is not in general the product of the partial traces. However, given the structure of the initial and final states in Eq.~(\ref{final}), and Eq.~(\ref{initial}), one can show directly that if $\Pi,\Pi'$ act as the identity on the ``2'' subsystems, then
\be
tr[\sigma(S) \Pi\rho\Pi']  = tr_1[\sigma_1(S) \Pi \rho_1 \Pi']
\ee
where 
\begin{align}
\sigma_1(S) &= tr_2\Big[\Big(|0\rangle\langle 0|_{M_2}\otimes 1\Big)\sigma(S)\Big] \\
\rho_1 &= tr_2[\rho].
\end{align}
It follows trivially that 
\be
tr[\sigma(S) \Pi\rho\Pi'] - tr[\Pi\rho\Pi'] = tr_1[(\sigma_1(S) - 1)\Pi\rho_1\Pi'].
\ee
If $\sigma_1$ is sufficiently close to the identity, this difference will be small. Since $\Pi,\Pi'$ are projection operators and $\rho_1$ is a normalized density matrix, the absolute value of the RHS can be no larger than the norm of the largest (in absolute value) eigenvalue of $(\sigma_1(S) - 1)$. In fact, it is easy to check that 
\be
\sigma_1(S) = \sigma_{M_1,in_1}(S)\otimes 1_{out_1}
\ee
so we can focus on bounding the eigenvalues of $(\sigma_{M_1,in_1} - 1)$. Explicitly, 
\be
\label{sigma}
\langle \ell'm'|\sigma_{M_1,in_1}(S)|\ell m\rangle = n\sum_k S_{0\ell,km}S^*_{0\ell',km'},
\ee
where the $\ell,\ell'$ indices label the $M_1$ system, and the $m,m'$ indices label the $in_1$ system. We are using a notation $S_{ij} = S_{ab,cd}$ for the $S$ matrix, in which the $a$ and $b$ indices represent the tensor decomposition of the left ($i$) index, while $c$ and $d$ represent the tensor decomposition of the right ($j$) index.

A simple but rather clumsy bound on the largest eigenvalue of $\sigma_{M_1,in_1}$ can be obtained from the square root of the sum of the squares of all the eigenvalues.\footnote{This is the most obvious place to try to improve the one-fifth bound.} We therefore have that
\begin{align}
\int dS \, \Big|tr[\sigma(S) \Pi\rho\Pi'] - tr[\Pi\rho\Pi']\Big| &\le \int dS \  \left(tr[(\sigma_{M_1,in_1}(S) - 1)^2]\right)^{1/2} \\
&\le \left(\int dS \  tr[(\sigma_{M_1,in_1}(S) - 1)^2]\right)^{1/2},
\end{align}
where the second line follows from the convexity of the square root. The expression inside the integral is now a quartic function of $S$, and its Haar average can be evaluated using the standard formulas\footnote{{For a general polynomial of $S$, the coefficients of the various $\delta$ function terms in the Haar integral are known as the Weingarten function.}}
\begin{align}
\int dS \ S_{ij}S^*_{i' j'} &= \frac{1}{N}\delta_{i i'}\delta_{j j'} \notag \\
\int dS \ S_{i_1 j_1}S_{i_2 j_2}S^*_{i_1' j_1'} S^*_{i_2' j_2'} &= \frac{1}{N^2-1}\Big(\delta_{i_1 i_1'}\delta_{i_2 i_2'}\delta_{j_1j_1'}\delta_{j_2j_2'} + \delta_{i_1 i_2'}\delta_{i_2 i_1'}\delta_{j_1j_2'}\delta_{j_2j_1'} \Big) \\ &\hspace{40pt} -\frac{1}{N(N^2-1)}\Big(\delta_{i_1i_1'}\delta_{i_2 i_2'}\delta_{j_1j_2'}\delta_{j_2j_1'} + \delta_{i_1 i_2'}\delta_{i_2 i_1'}\delta_{j_1j_1'}\delta_{j_2j_2'}  \Big).\notag
\end{align}
The terms on the second line are subleading, and can be safely ignored in this calculation. Substituting in Eq.~(\ref{sigma}) and contracting indices appropriately, one finds Eq.~(\ref{answer}).

\end{appendix}

\end{document}